# Bifacial Si Heterojunction-Perovskite Organic-Inorganic Tandem to Produce Highly Efficient Solar Cell

Reza Asadpour[1*], Raghu V. K. Chavali[1*], M. Ryyan Khan[1*], and Muhammad A. Alam[1, a)]

[1]*Electrical and Computer Engineering Department, Purdue University, West Lafayette, IN-47907, USA*

As single junction thin-film technologies, both Si heterojunction (HIT) and Perovskite based solar cells promise high efficiencies at low cost. One expects that a tandem cell design with these cells connected in series will improve the efficiency further. Using a self-consistent numerical modeling of optical and transport characteristics, however, we find that a traditional series connected tandem design suffers from low $J_{SC}$ due to band-gap mismatch and current matching constraints. It requires careful thickness optimization of Perovskite to achieve any noticeable efficiency gain. Specifically, a traditional tandem cell with state-of-the-art HIT ($\eta = 24\%$) and Perovskite ($\eta = 20\%$) sub-cells provides only a modest tandem efficiency of $\eta_T \sim 25\%$. Instead, we demonstrate that a bifacial HIT/Perovskite tandem design decouples the optoelectronic constraints and provides an innovative path for extraordinary efficiencies. In the bifacial configuration, the same state-of the-art sub-cells achieve a normalized output of $\eta_T^*=33\%$, exceeding the bifacial HIT performance at practical albedo reflections. Unlike the traditional design, this bifacial design is relatively insensitive to Perovskite thickness variations, which may translate to simpler manufacture and higher yield.

*Background and Motivation*. In recent years, the search for low-cost highly efficient solar cells has resulted in two cell technologies (inorganic a-Si/c-Si Heterojunction and organic Perovskite) with comparable, but complementary characteristics. On one hand, the standard a-Si/c-Si Heterojunction (HIT) technology is poised to capture a significant proportion of traditional c-Si solar cells due to its high efficiency and better temperature coefficients.[1,2] On the other hand, the recent dramatic gain in efficiency of Perovskite cells promises to finally realize the presumed cost advantages of organic solar cells.[3–5] With significant efforts in device analysis[6–11] and optimization[12–15], several groups have achieved efficiency $\eta \geq 22\%$ HIT cells.[16] In the case of Perovskite based solar cells, with improvements in processing and material quality, the state-of-the-art cells reached an efficiency of ~20%.[17–19]

A Perovskite/HIT tandem design that can suppress individual bottlenecks and take advantage of their complementary characteristics may improve the efficiency further. Recently, the efficiency gain of tandem designs with Perovskite as the top cell and a range of bottom cell have been explored. Using a four terminal configuration, Bailie *et al.*[20] obtained a 17% and 18.6% efficient tandem cell with mc-Si ($\eta \sim 11\%$) and copper indium gallium selenide (CIGS, $\eta \sim 17\%$) bottom cells, respectively. Similarly, Löper *et al.*[21] obtained a 13.4% efficient tandem cell with presumably a highly efficient a-Si:H/c-Si heterojunction bottom cell using the same configuration and Mailoa *et al.*[22] obtained a 13.7% efficient cell with c-Si bottom cell using the two terminal configuration. These studies indicate that a poor performing top cell can significantly limit the performance of these traditional tandem cell designs. The tandem efficiency improvement may saturate for the state-of-art sub-cells.

There has been some efforts to predict the theoretical limits for these traditional tandem designs using Perovskite sub-cell on top of c-Si [23] or a-Si/c-Si heterojunction bottom cell.[24] These studies are generally not self-consistent; they rely only on detailed optical modeling. The conclusions based on carrier transport modelled by simple one-diode compact model may not be definitive, at least in the case of a-Si/c-Si heterojunction solar cells.[8]

From the above discussion, we observe that there has been some efforts to push the practical conversion efficiency using the traditional tandem designs. However, the gain using traditional tandem designs is not significant and theoretical limits based on full self-optoelectronic simulation is still lacking. In this letter, using detailed optical and carrier transport modeling, first, we re-explore the traditional tandem cell design. Starting from state-of-the-art cell parameters for the sub-cells (~20% efficient Perovskite cell, ~24% efficient HIT cell), an optimized tandem cell offers only a modest improvement, $\eta_T \sim 25\%$. Moreover, the optimization dictates a strict control over the Perovskite layer thickness —even a ~20nm deviation from the optimum thickness would degrade the net efficiency by 1%.

Instead, we suggest that another design involving a HIT and Perovskite *bifacial* tandem cells can drastically circumvent the constraints of the classical design, and increase the *normalized* output to $\eta_T^* \sim 33\%$ (output power normalized to 1-sun). Indeed, the bifacial tandem would outperform the bifacial HIT cell for typical albedo reflection ($R_A < 40\%$). Further, the bifacial tandem design would be insensitive to the sub-cell thickness variation which obviates the need for layer optimization. The output gain is not restricted to state-of-the-art champion cells; even sub-optimal Perovskite cell and standard HIT cell would benefit

---

a) Electronic mail: alam@purdue.edu

* Equal Contribution

substantially from this strategy.[25]

A typical Perovskite solar cell[3,26] consists of PEDOT:PSS as the hole transport material (HTM) and PCBM as the electron transport material (ETM) on either sides of the absorber layer. This Perovskite sub-structure is then contacted by ITO and Al. This configuration is shown in Fig. 1(b) without the Al back contact. The HIT cell uses c-Si as the active material with $p^+$ and $n^+$ doped a-Si passivation layers which act as the emitter (i.e., HTM) and the back surface field (ETM), respectively. The cell is contacted with ITO at the front and ITO/Al at the back.

The traditional tandem cell consists of higher bandgap Perovskite ($CH_3NH_3PbI_3$, $E_G^{PVK} \approx 1.55eV$) sub-structure stacked on top of the lower bandgap ($E_G^{cSi} \approx 1.12eV$) c-Si HIT-subcell (see Fig. 1(b)).

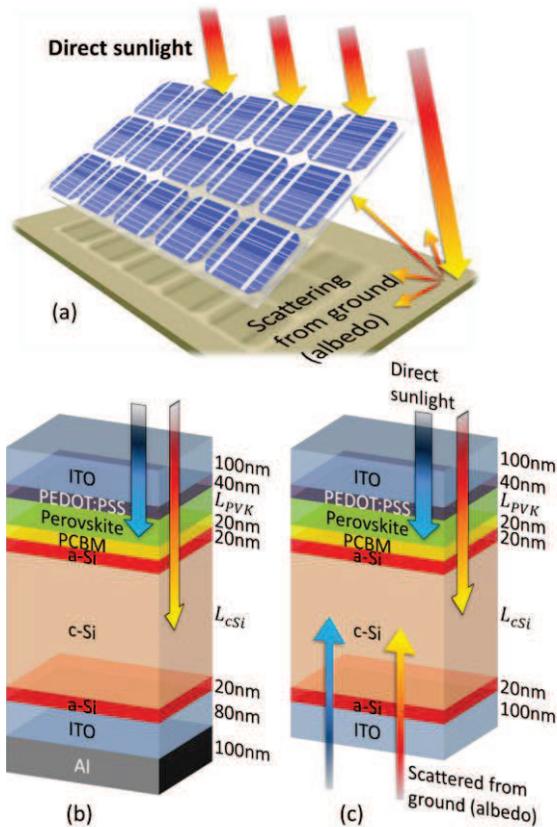

Fig. 1: (a) A solar panel generally utilizes the direct illumination of sunlight. A fraction of the light can also be scattered from the ground onto the back of the panel. (b) Traditional tandem structure, and (c) a bifacial tandem structure.

The bifacial tandem cell has a similar design except for the absence of Al back contact. This allows the absorption of reflected light, as shown in Fig. 1(c). We will consider both classical and bifacial tandem designs, to demonstrate why and how bifacial tandem outperforms the classical design.

*Simulation Framework.* The absorption spectrum and spatial photo-generation profile in the layered solar cell structure is calculated by solving the Maxwell's equations using Transfer Matrix Method (TMM)[27] under illumination of AM 1.5G solar spectrum over the wavelength range 300–1500nm. The materials in different layers of the cell are characterized by experimentally measured absorption coefficient and refractive indices reported in the literature [25,28–30].

The transport of charged carriers (electrons and holes) is modeled by generalized drift-diffusion formalism[31] using the carrier generation profile found from the optical simulation. The carrier transport in the cell is simulated by a self-consistent solution of Poisson and continuity equations by MEIDCI$^{TM}$ device simulator.[29] The contacts are presumed to be ohmic. See Table S1 for model equations, and Table S2 and Table S3 for simulation parameters in SI.[25] The parameters are consistent with literature yielding in experimentally observed device behaviors.

To model the tandem cell designs, first, we perform optical simulation of the full structure. Then, the carrier generation profiles so obtained are used in the transport simulator to separately characterize the $J-V$ for sub-cells. Finally, the sub-cells are connected in series using a circuit model (with negligible series resistance) to obtain the tandem characteristics.

*Traditional Tandem Configuration.* The electronic properties of the materials (Perovskite, a-Si, and c-Si) considered in this study are chosen in such a way to obtain the state-of-the-art efficiencies of 24% for HIT cell and 20% for the Perovskite cell. (see Table S2 and Table S3 in SI for parameter list).[25] The corresponding $J-V$ characteristics are provided in Fig. 3(a). In the following discussions, we will show how these two technologies can be arranged in tandem to achieve high performance solar cells.

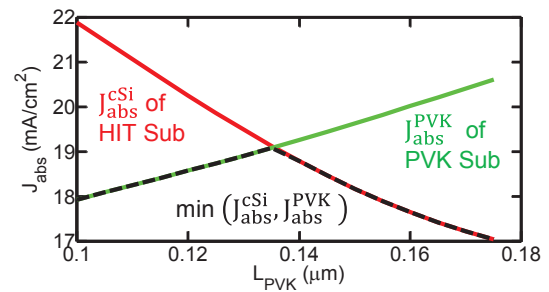

Fig. 2: Absorbed sunlight (in terms of corresponding current density $J_{abs}$) is shown for Perovskite (green) and HIT (red) subcells as a function of $L_{PVK}$. The short circuit current of the series tandem will be limited to $\min(J_{abs}^{cSi}, J_{abs}^{PVK})$ as shown by the dashed black lines. We set $L_{cSi} = 200\mu m$.

Perovskite and HIT sub-cells are connected in series tandem configuration; see Fig. 1(b). Ideally, a top cell with $E_G \sim 1.7eV$ is required to match current produced by c-Si bottom sub-cell, to achieve the maximum efficiency for the combination.[33] Here, a sufficiently thick top sub-cell should absorb *all the photons* above its bandgap, transmitting the rest of the solar spectrum to the bottom cell. The bottom sub-cell then absorbs above its bandgap from the remaining part of the spectrum (see absorption spectrum in Fig. S2 in SI).[25] The optimum bandgap for the top cell is selected such that, the sub-cell currents are matched and output power is maximized under the above mentioned operation principles.

Unfortunately, Perovskite bandgap (1.55eV) is considerably smaller than the optimum bandgap of the top-cell ($E_G \sim 1.7 eV$).[33] This makes the traditional tandem design sub-optimal, because 300-400nm thick typical Perovskite top-cell would absorb so many photons that the bottom c-Si cell would not be able to produce sufficient $J_{sc}$ — this mismatch in $J_{SC}$ dramatically suppresses the power output from the tandem cell. Even for a relatively thin top-cell (e.g., $L_{PVK} \sim 170$nm, see Fig. 2(b)), the currents are mismatched by 4 mA (red line, $J_{abs}^{cSi}$; green line, $J_{abs}^{PVK}$). Therefore, delicate thickness control is *essential* in designing traditional tandem cell with these mismatched sub-cells.

Fig. 2 shows that the sub-cell currents can be matched by varying the thickness of Perovskite ($L_{PVK}$) layer. The minimum absorption between the two sub-cells approximately defines the tandem $J_{SC}$. Therefore, we maximize $\min(J_{abs}^{PVK}, J_{abs}^{cSi})$ for achieving the highest overall current (black dashed line in Fig. 2). As the sensitivity of $J_{SC}$ ($\sim \min(J_{abs}^{PVK}, J_{abs}^{cSi})$) on the c-Si layer thickness is negligible, $L_{cSi} = 200$μm in HIT cell is chosen for the optimized tandem cell. In contrast, the matched current is highly sensitive to $L_{PVK}$ and even 20nm variation will lead to more than 1mA/cm$^2$ lowering of short circuit current—this translates to more than 1% loss in tandem cell efficiency. The optical simulation indicates that the efficiency would be optimal at $L_{PVK} = 135$nm.

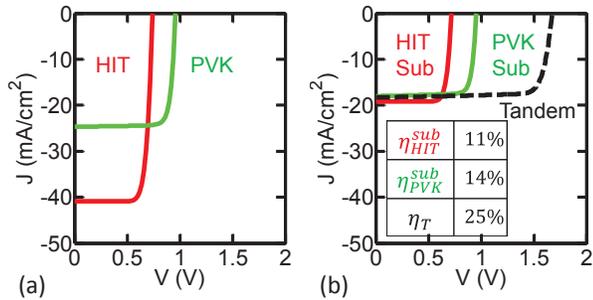

Fig. 3: $J - V$ characteristics of (a) individual HIT and Perovskite cells (b) HIT (red) and Perovskite (green) sub-cell $J - V$ characteristics are shown along with the tandem cell (black dashed line). Inset table shows sub-cells contribution to the efficiency.

Once the cell thicknesses are determined from optical simulation, the analysis of carrier transport produces the full $J - V$ characteristics. First, consider the individual $J - V$ characteristics of Perovskite and HIT cells shown in Fig. 3(a). HIT solar cell has a higher $J_{SC}$ but lower $V_{OC}$ compared to Perovskite cell (as $E_G^{cSi} < E_G^{PVK}$). In a tandem structure, the sub-cell currents must be matched, as in Fig. 3(b). The $J_{SC}$ in the tandem cell (~18mA/cm$^2$) after current matching is lower than both the Perovskite (~24.5mA/cm$^2$) and HIT (~40.5mA/cm$^2$) cells. The $V_{OC}$ in the tandem cell adds up to ~1.65V from the sub-cells. The total efficiency is $\eta_T \sim 25\%$, which is only slightly higher than the individual cells, $\eta_{HIT} \sim 24\%$ and $\eta_{PVK} \sim 20\%$. Therefore, while with *careful layer optimization*, it is possible to obtain modest efficiency gains through traditional tandem configurations, but it may not be cost effective. Note that, a part of the efficiency gain arises from not having ITO between the two sub-cells, which reduces parasitic absorption[21], and offers better impedance matching for light between the top and the bottom cells.

***Bifacial Tandem Configuration.*** The efficiency ($\eta_T$) gain in traditional tandem configuration, unfortunately, is highly sensitive to $L_{PVK}$; any deviation from the optimum (Fig. 2, 135nm) erodes the gain considerably. In practice, it is difficult to control $L_{PVK}$ within 20-30nm by spin or drop-casting, so that average gain would be lower than predicted by the optimal. Also, in a tandem, the sub-cells underperform compared to their individual limits ($\eta_{PVK}^{sub} \sim 14\%$ vs. $\eta_{PVK} \sim 20\%$; $\eta_{HIT}^{sub} \sim 11\%$ vs. $\eta_{HIT} \sim 24\%$). A reduced coupling between the sub-cells is desired and the characteristic bifacial configuration of the HIT cell (see Fig. 1(c)) offers a simple solution, with $\eta_T^* \to 33\%$, as discussed below.

The bifacial design allows light to enter the device from both top and bottom faces. In a HIT cell, the back ITO must be inserted between a-Si and metal contact to improve the optical properties and reliability,[34] so that a transparent back-contact is available at no extra cost. Fig. 1(a) shows that the bifacial panel accepts direct solar illumination from the front, and albedo reflected light from the background (grass, concrete, snow, etc.). The traditional and the bifacial

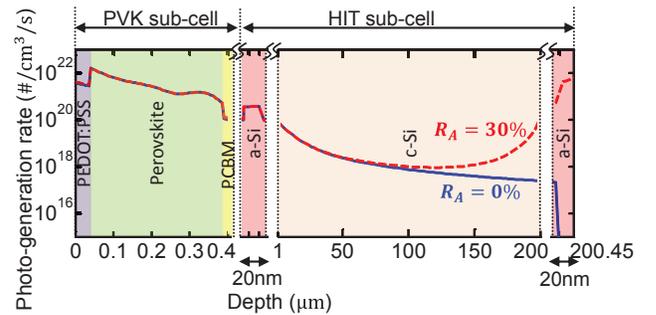

Fig. 4: Spatial photo-generation profile in the tandem cell. The primary sunlight is from the left.

Perovskite-HIT tandem are identical (Fig. 1(b) vs. 1(c)), except for the back Al contact. As a result, slight increase in series resistance is expected for the bifacial design; for this study, we will assume that this effect is negligible.

The front side of the cell is exposed to direct sunlight—the high energy part of the spectrum for $> E_G^{PVK}$ (1.55eV) is now mostly absorbed in a 350nm Perovskite sub-cell (see absorption spectrum in Fig. S3 in SI).[25] The bottom HIT sub-cell absorbs few photons from the remaining spectrum, the blue line in Fig. 4. In a typical tandem, with no albedo reflectance ($R_A = 0$), $\eta_T^*$ would be severely affected. However, the partially reflected light from the ground/surroundings, characterized by albedo reflectance ($R_A < 100\%$), improves photo-generation in the HIT sub-cell considerably (red dashed line in Fig. 4). For the traditional tandem cell discussed earlier, $\eta_T$ was limited by the constraint of matched sub-cell currents. Since bifacial tandem improves the $J_{SC}^{HIT}$, one can use a thicker Perovskite to improve $J_{SC}^{PVK}$, free from typical constraints of tandem cell, i.e., restriction on $L_{PVK}$ vs. $L_{cSi}$. For this analysis, we choose these thicknesses to be 350nm and 200μm respectively.

We can now compare the bifacial HIT cell and the bifacial tandem for varying $R_A$. For low $R_A$ ($< 20\%$), the Perovskite has higher current than the HIT sub-cell (see Fig. 5(a)), and the efficiency is limited by the HIT cell. As we increase scattered light entering through the bottom cell, the absorption in the HIT sub-cell will increase linearly with $R_A$. However, the tandem $J_{SC}$ will be limited by the lower of the two sub-cells currents, shown by the black line in Fig. 5(a). Beyond $R_A > 20\%$, the tandem current is limited by the Perovskite sub-cell. This also explains why the tandem cell output $\eta_T^*$ (output power normalized to 1-sun illumination) increases with $R_A$ and then saturates to ~33% after $R_A > 20\%$ (see Fig. 5(b)). Further, from Fig. 5(b) we observe that for a practical range of $R_A < 40\%$ the tandem design (black line) outperforms the typical bifacial HIT cell (blue solid line) by a considerable margin. Obviously, this bifacial design requires a slightly modified stacking of the cells– a suggested process flow is described in SI.[25]

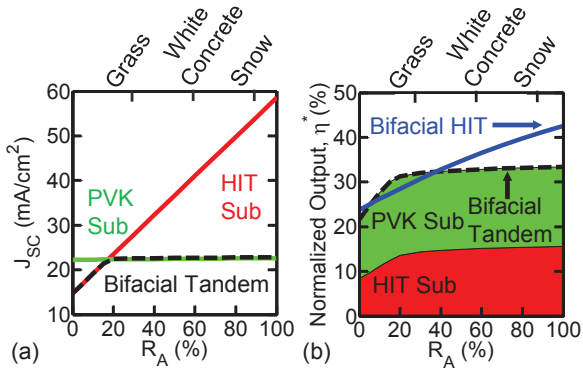

Fig. 5: (a) The $J_{SC}$ of the sub-cells and bifacial tandem as function of albedo $R_A$. (b) Output $\eta_T^*$ (normalized to 1-sun) of the Perovskite and HIT sub-cells (green and red areas) are shown. The bifacial tandem (black dashed line) outperforms the bifacial HIT cell (blue solid line) for $R_A < 40\%$. Common backgrounds such as grass, white concrete, and snow have $R_A$ of 30%, 60%, and 90%, respectively[35] (some variations in these $R_A$ values are observed in literature[36]).

*Summary.* In this paper, we have explored the performance potential of Perovskite-HIT tandem cell based on state-of-the-art sub-cells. We find that a traditional tandem design requires an optimized Perovskite thickness of $L_{PVK} \sim 135nm$ to provide a modest 25% efficient cell. *Unfortunately*, the efficiency gain is compromised due to sensitivity to $L_{PVK}$ and both the cells underperform compared to their individual efficiencies. Through a bifacial tandem design one can resolve the current matching problem and improve the performance to 33%. As an added advantage, we observe that this efficiency gain is insensitive to thicknesses of Perovskite and c-Si layers. Further, it outperforms the bifacial HIT cell over a practical range of albedo reflection. Therefore, it offers a viable, robust HIT-Perovskite tandem for low-cost, highly-efficient PV technology.

[1] K. Maki, D. Fujishima, H. Inoue, Y. Tsunomura, T. Asaumi, S. Taira, T. Kinoshita, M. Taguchi, H. Sakata, H. Kanno, and E. Maruyama, in *2011 37th IEEE Photovolt. Spec. Conf.* (IEEE, 2011), pp. 000057–000061.
[2] M. Taguchi, A. Yano, S. Tohoda, K. Matsuyama, Y. Nakamura, T. Nishiwaki, K. Fujita, and E. Maruyama, IEEE J. Photovolt. **4**, 96 (2014).
[3] J. You, Z. Hong, Y. (Michael) Yang, Q. Chen, M. Cai, T.-B. Song, C.-C. Chen, S. Lu, Y. Liu, H. Zhou, and Y. Yang, ACS Nano **8**, 1674 (2014).
[4] P. Docampo, J.M. Ball, M. Darwich, G.E. Eperon, and H.J. Snaith, Nat Commun **4**, (2013).
[5] P.P. Boix, K. Nonomura, N. Mathews, and S.G. Mhaisalkar, Mater. Today **17**, 16 (2014).
[6] a. S. Gudovskikh and J.P. Kleider, Appl. Phys. Lett. **90**, 034104 (2007).
[7] R.S. Crandall, E. Iwaniczko, J. V Li, and M.R. Page, J. Appl. Phys. **112**, 093713 (2012).
[8] R.V.K. Chavali, J.R. Wilcox, B. Ray, J.L. Gray, and M.A. Alam, IEEE J. Photovolt. **4**, 763 (2014).
[9] R. Varache, J.P. Kleider, W. Favre, and L. Korte, J. Appl. Phys. **112**, (2012).
[10] S. De Wolf and M. Kondo, Appl. Phys. Lett. **90**, 042111 (2007).
[11] R.V.K. Chavali, S. Khatavkar, C. V Kannan, V. Kumar, J.L. Gray, and M.A. Alam, IEEE J. Photovolt. (2015).
[12] L. Korte, E. Conrad, H. Angermann, R. Stangl, and M. Schmidt, Sol. Energy Mater. Sol. Cells **93**, 905 (2009).
[13] H. Angermann, L. Korte, J. Rappich, E. Conrad, I. Sieber, M. Schmidt, K. Hübener, and J. Hauschild, Thin Solid Films **516**, 6775 (2008).
[14] a. Descoeudres, L. Barraud, S. De Wolf, B. Strahm, D. Lachenal, C. Guérin, Z.C. Holman, F. Zicarelli, B. Demaurex, J. Seif, J. Holovsky, and C. Ballif, Appl. Phys. Lett. **99**, 123506 (2011).
[15] Z.C. Holman, A. Descoeudres, L. Barraud, F.Z. Fernandez, J.P. Seif, S. De Wolf, and C. Ballif, IEEE J. Photovolt. **2**, 7 (2012).
[16] C. Ballif, S. De Wolf, A. Descoeudres, and Z.C. Holman, *Advances in Photovoltaics: Part 3* (Elsevier, 2014), pp. 73–120.
[17] H. Zhou, Q. Chen, G. Li, S. Luo, T. Song, H.-S. Duan, Z. Hong, J. You, Y. Liu, and Y. Yang, Science **345**, 542 (2014).
[18] M.A. Green, K. Emery, Y. Hishikawa, W. Warta, and E.D. Dunlop, Prog. Photovolt. Res. Appl. **23**, 1 (2015).
[19] W. Nie, H. Tsai, R. Asadpour, J.-C. Blancon, A.J. Neukirch, G. Gupta, J.J. Crochet, M. Chhowalla, S. Tretiak, M.A. Alam, H.-L. Wang, and A.D. Mohite, Science **347**, 522 (2015).

[20] C.D. Bailie, M.G. Christoforo, J.P. Mailoa, A.R. Bowring, E.L. Unger, W.H. Nguyen, J. Burschka, N. Pellet, J.Z. Lee, M. Grätzel, R. Noufi, T. Buonassisi, A. Salleo, and M.D. McGehee, Energy Environ. Sci. **8**, 956 (2015).

[21] P. Löper, S.-J. Moon, S. Martín de Nicolas, B. Niesen, M. Ledinsky, S. Nicolay, J. Bailat, J.-H. Yum, S. De Wolf, and C. Ballif, Phys Chem Chem Phys (2014).

[22] J.P. Mailoa, C.D. Bailie, E.C. Johlin, E.T. Hoke, A.J. Akey, W.H. Nguyen, M.D. McGehee, and T. Buonassisi, Appl. Phys. Lett. **106**, 121105 (2015).

[23] B.W. Schneider, N.N. Lal, S. Baker-Finch, and T.P. White, Opt. Express **22**, A1422 (2014).

[24] M. Filipič, P. Löper, B. Niesen, S. De Wolf, J. Krč, C. Ballif, and M. Topič, Opt. Express **23**, A263 (2015).

[25] See supplemental material at [URL will be inserted by AIP] for Perovskite (n, k) Fig S1, absorption spectrum Fig S2-S4, moderate cell/sub-cell results Fig S5-S8, energy band diagrams Fig S9-S10, transport equations and parameters Table S1-S3.

[26] J. Seo, S. Park, Y. Chan Kim, N.J. Jeon, J.H. Noh, S.C. Yoon, and S.I. Seok, Energy Env. Sci **7**, 2642 (2014).

[27] L.A.A. Pettersson, L.S. Roman, and O. Inganäs, J. Appl. Phys. **86**, 487 (1999).

[28] V.S. Gevaerts, L.J.A. Koster, M.M. Wienk, and R.A.J. Janssen, ACS Appl. Mater. Interfaces **3**, 3252 (2011).

[29] H. Hoppe, N.S. Sariciftci, and D. Meissner, Mol. Cryst. Liq. Cryst. **385**, 113 (2002).

[30] G. Xing, N. Mathews, S.S. Lim, N. Yantara, X. Liu, D. Sabba, M. Grätzel, S. Mhaisalkar, and T.C. Sum, Nat. Mater. **13**, 476 (2014).

[31] R.F. Pierret, *Advanced Semiconductor Fundamentals*, 2 edition (Prentice Hall, Upper Saddle River, N.J, 2002).

[32] *Taurus Medici* (Synopsys, United States, 2010).

[33] S.P. Bremner, M.Y. Levy, and C.B. Honsberg, Prog. Photovolt. Res. Appl. **16**, 225 (2008).

[34] a. Banerjee and S. Guha, J. Appl. Phys. **69**, 1030 (1991).

[35] Prism Solar Technologies, (n.d.).

[36] U. Feister and R. Grewe, Photochem. Photobiol. **62**, 736 (1995).

# SUPPLEMENTARY INFORMATION

# Bifacial Si Heterojunction-Perovskite Organic-Inorganic Tandem to Produce Highly Efficient Solar Cell

Reza Asadpour[1*], Raghu V. K. Chavali[1*], M. Ryyan Khan[1*], and Muhammad A. Alam[1, a)]

[1]Electrical and Computer Engineering Department, Purdue University, West Lafayette, IN-47907, USA. Author to whom correspondence should be addressed. Electronic mail: alam@purdue.edu
(* Equal Contribution)

## Absorption spectrums:

There are a number of different measurements in the literature regarding the optical properties (complex refractive index) for Perovskite[1–4] and we used the one from Xing *et al.*[1] The extinction coefficient has been set to zero below the band-edge ($\lambda > 800\ nm$ see Fig. S1)—this yields an abrupt jump in the simulated absorption spectrums (discussed below). To understand the bifacial tandem, we will present two absorption spectrums: (i) for direct sunlight illumination form the top, and (ii) for scattered light incident from the back of the device.

Fig. S2 shows the absorption spectrum for the optimized tandem structure. Most of the high energy photons ($\lambda < 500$nm) are absorbed in the Perovskite sub-cell. This suppresses parasitic absorption in a-Si layer, a beneficial effect of the tandem configuration. Similarly, for the bifacial configuration, the thick Perovskite layer absorbs almost all the photons above its band-edge, as shown in Fig. S3. This eliminates the possibility of parasitic absorption in a-Si.

For simulation of the light incident from the bottom of the cell (in the bifacial configuration), we assumed perfect impedance matching from air into the a-Si layer (the ITO has been neglected in this case). Such matching can be achieved by anti-reflection coatings or by textured surfaces. In this case, as shown in Fig. S4, there is significant parasitic absorption in back a-Si layer. The rest of the photons are absorbed in c-Si (above the band-edge). Since c-Si is a smaller bandgap material, there is no absorption in Perovskite layer when light is incident from the back surface.

The absorption spectrums for light incident from the top and from albedo reflection are calculated as a function of the device depth ($x$), represented by $A_{top}(\lambda, x)$ and $A_{bot}(\lambda, x)$ respectively. We then find the carrier generation profile in the tandem device:

$$n_{Gen}(x) = \int \left( \frac{I_0(\lambda)}{hc/\lambda} A_{top}(\lambda, x) + R_A \times \frac{I_0(\lambda)}{hc/\lambda} A_{bot}(\lambda, x) \right) d\lambda$$

Here, $I_0(\lambda)$ is the AM1.5 spectrum, and $R_A$ is the albedo reflectance. The number of photons (#/wavelength/area) at $\lambda$ is given by $I_0/(hc/\lambda)$.

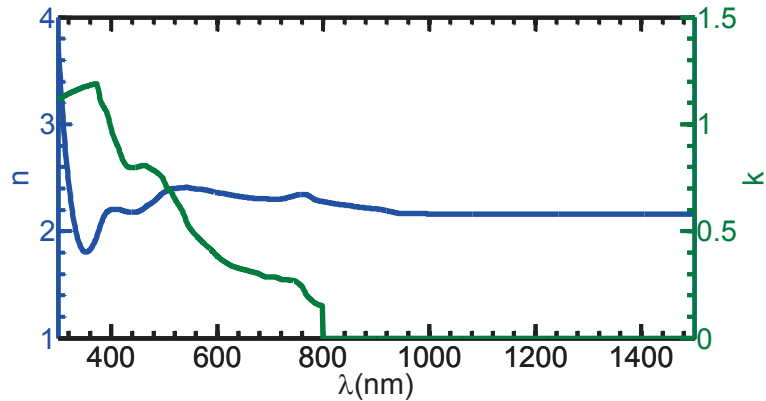

**Fig. S1** Refractive index of Perovskite reproduced from[1]. For $\lambda > 800\ nm$ the extinction coefficient (k) has been set to zero.

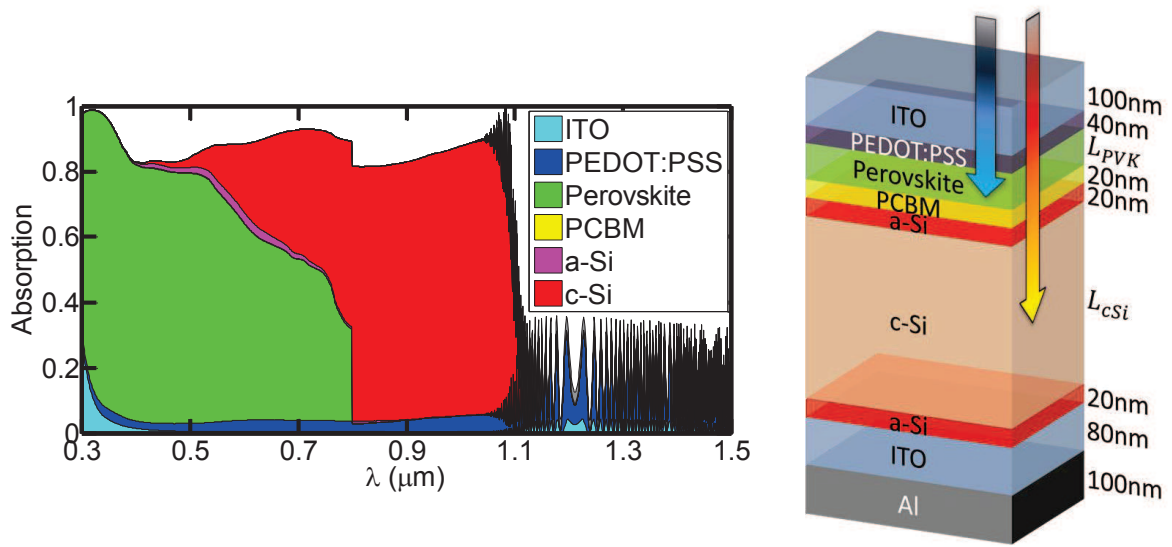

**Fig. S2** Absorptance spectrum of optimized thickness classical tandem solar cell.

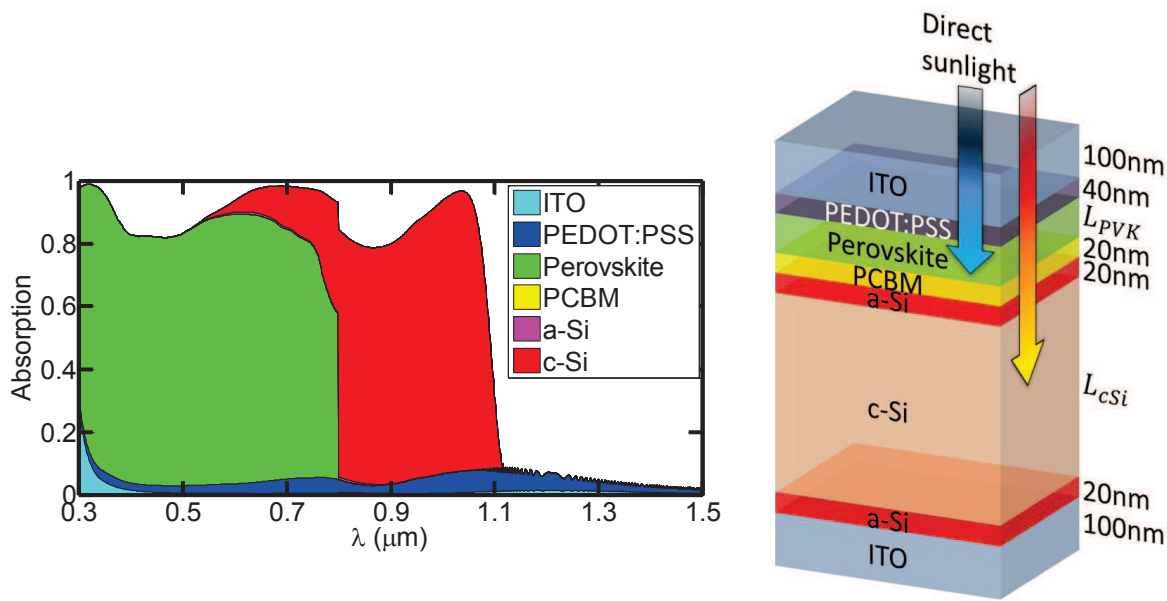

**Fig. S3** Absorption spectrum of *top incident light* for bifacial tandem solar cell.

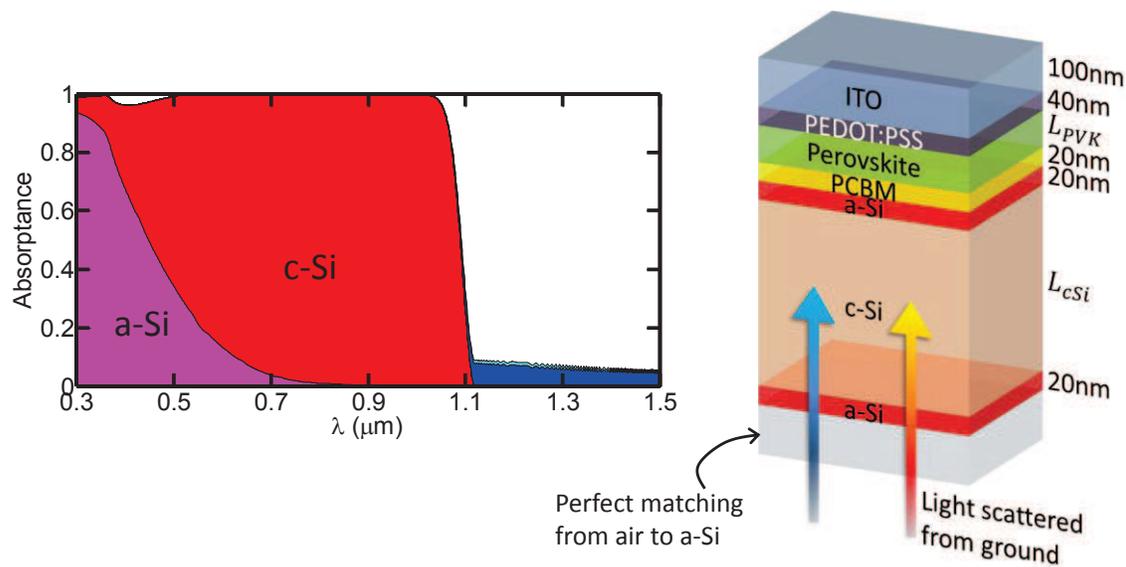

**Fig. S4** Absorption spectrum *of bottom incident light* (Albedo reflection) for bifacial tandem solar cell.

## Moderate efficiency sub-cell results:

In the HIT or Perovskite cell technology, most of the cells do not reach the champion cell efficiencies. Typical cells have somewhat lower efficiency. For example, we may expect moderately performing HIT and Perovskite cells to reach ~20% and ~17.5%, respectively. It would be interesting to see if the advantages of bifacial design (over traditional design) persist for these cells as well.

The typical HIT cell (~20% efficiency) was realized by using doping level of $1 \times 10^{19}\ /cm^3$ for n-layer and $1 \times 10^{16}\ /cm^3$ for p-layer also decreasing the lifetime from $10\ ms$ to $0.5\ ms$ in c-Si and reducing hole mobility by a factor of 2 in both n-layer and p-layer. For the perovskite cell (~17.5% efficiency) mobility was decreased from $40\ V.s/cm^2$ to $30\ V.s/cm^2$ and lifetime from $20\ ns$ to $3.3\ ns$ in Perovskite layer. The tandem cell that combines both has 23%. (See Fig. S5**Error! Reference source not found.**) The relative improvement is more pronounced for the case of moderate 20% HIT cell that has been enhanced to 23% in tandem structure.

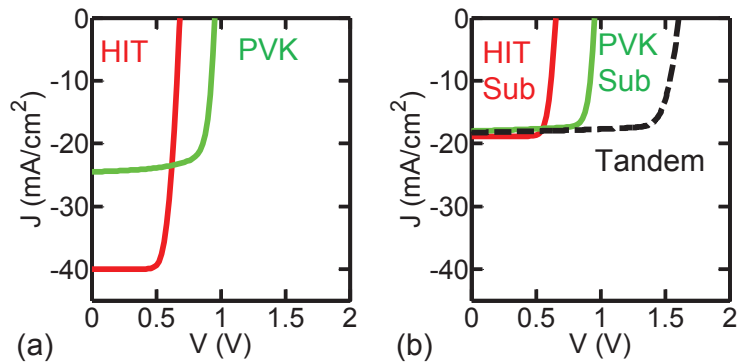

**Fig. S5** (a) Typical separate HIT and PVK cell and (b) Typical sub-cells and tandem J-V characteristics.

As can be seen from Fig. S6**Error! Reference source not found.** the overall trend of tandem bifacial outperforming the HIT bifacial cell for up to 50% albedo reflection remains the same. The bifacial tandem can go up to 29% normalized output.

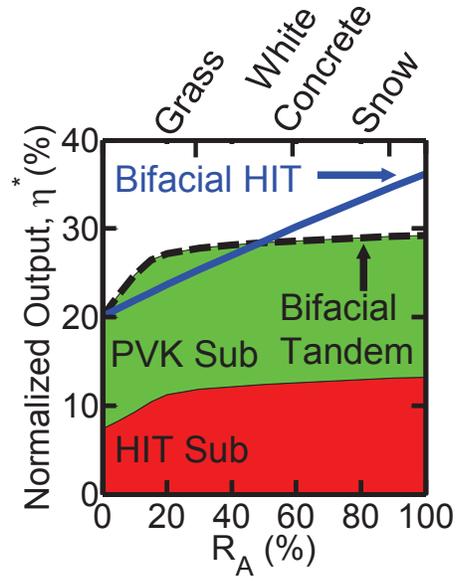

**Fig. S6** Comparison of normalized efficiency of bifacial HIT cell and tandem cell made of typical cells. Bifacial tandem cell outperforms the bifacial HIT structure up to 50% of albedo reflection.

## Thickness study for individual cells:

Fig. S7 shows thickness dependence of typical Perovskite solar cell characteristics. The efficiency overall follows the $J_{SC}$ trend. The increase in the current density for the thickness less than 0.12 $\mu m$ is due to interference of back mirror (contact).

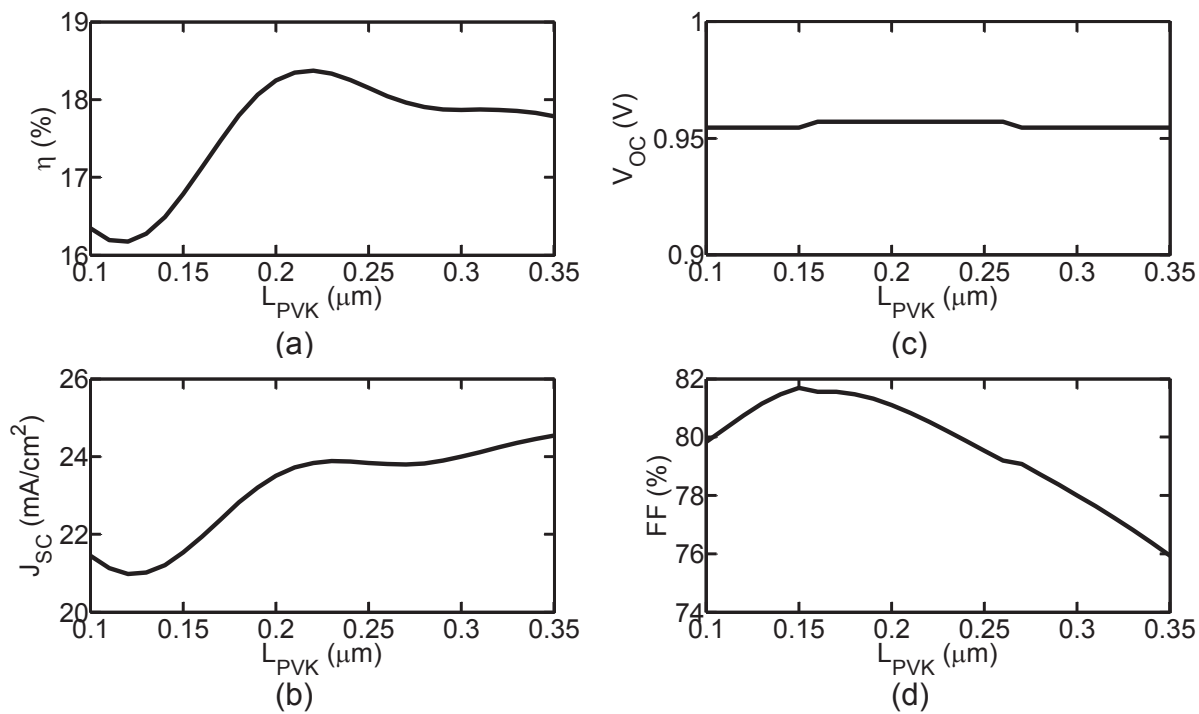

**Fig. S7** Thickness dependence of Perovskite cell characteristics.

HIT cell shows robustness with respect to thickness for a wide range. (See Fig. S8) Therefore in the optimization of tandem, we only considered the variation of the perovskite layer thickness (see main manuscript).

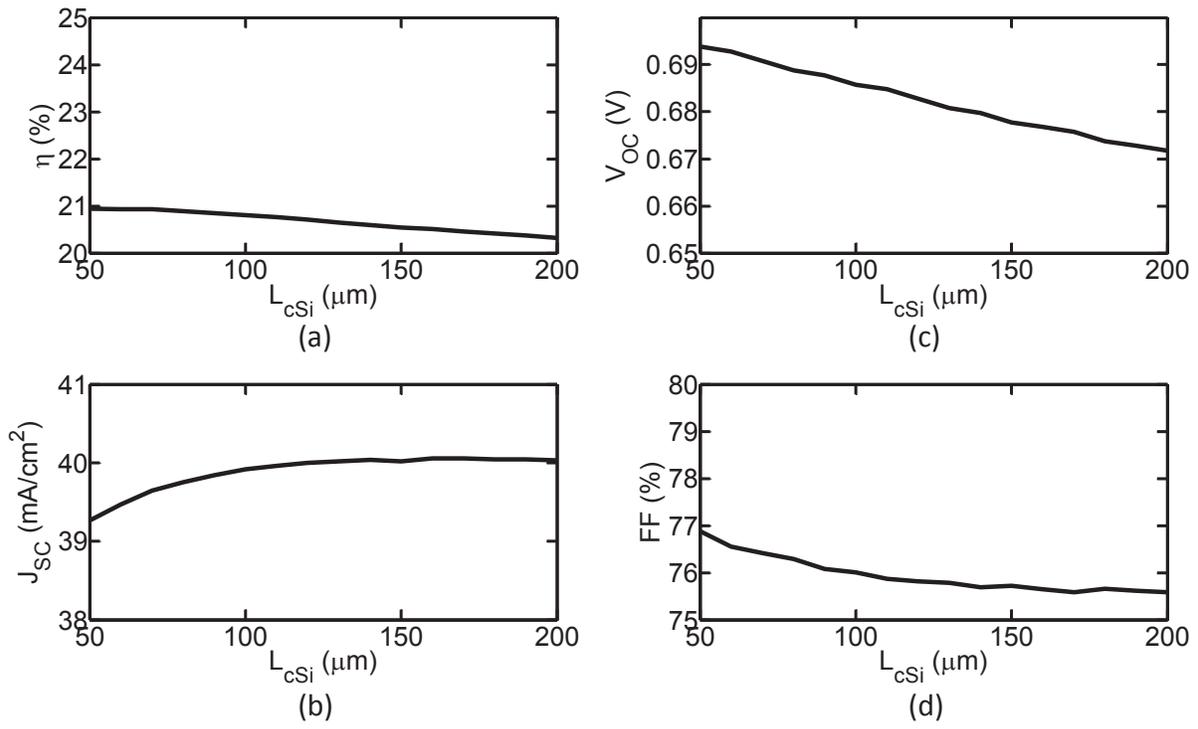

**Fig. S8** Thickness dependence of HIT cell characteristics.

## Transport simulation:

Table S1 to Table S3 summarize the equations and parameters used for transport simulation of the solar cells including the sub-cells, individual cells, and bifacial HIT structure. The parameters used for Perovskite cell and sub-cells are set to fit the experimental data from Nie et al.[5]

**Table S1.** Equations for carrier transport

Poisson Equation:

$$\epsilon_r \epsilon_0 \nabla^2 \psi = -q\,(n_h - n_e)$$

Continuity:

$$\nabla J_{e,h} = \left(G_{e,h} - R_{e,h}(n_e, n_h)\right)$$

Drift-Diffusion:

$$J_{e,h} = \mu_{e,h} n_{e,h}(-\nabla \psi) \pm D_{e,h} \nabla n_{e,h}$$

Recombination:

$$R_{e,h}(n_e, n_h) = B\left(n_e n_h - n_i^2\right) + \frac{n_e n_h - n_i^2}{\tau(n_e + n_h)}$$

Here, $\epsilon_r$ is the relative dielectric constant of the material, $n_{e,h}$ is the electron/hole concentration, $G_{e,h}$ is the generation profile coming from optical simulation, $R_{e,h}$ is the recombination term including direct and Shockley-Read-Hall recombination in case of Perovskite cell and also Auger recombination in case of HIT cell, $\mu_{e,h}$ is the electron/hole mobility, $D_{e,h}$ is the electron/hole diffusion coefficient, $n_i$ is the intrinsic carrier concentration, $B$ is the direct recombination coefficient, and $\tau$ is electron/hole lifetime.

The energy band diagram of the Perovskite sub-cell in short circuit condition is shown in Fig. S9. The Perovskite layer is fully depleted (from 0.040 $\mu m$ to 0.175 $\mu m$). Due to the electrical field in the depletion region, the generated charged carriers will be directed to corresponding contacts. PEDOT:PSS layer (from 0 $\mu m$ to 0.040 $\mu m$) acts as Hole Transport Material (HTM). PCBM layer (from 0.175 $\mu m$ to 0.195 $\mu m$) is acting as an Electron Transport Material (ETM) and blocking the holes to be collected in the wrong contact.

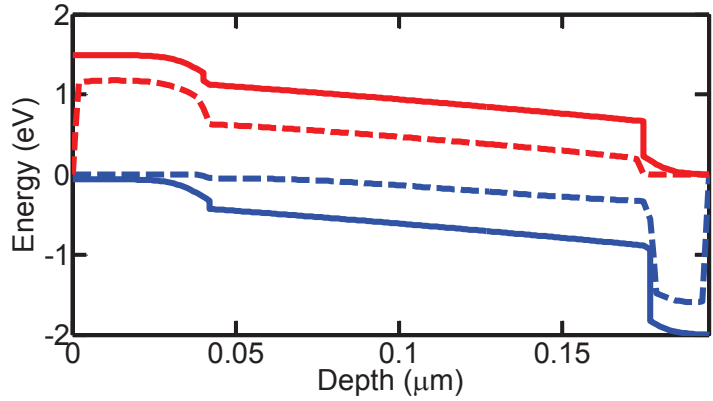

**Fig. S9** The energy band diagram of the Perovskite (PVK) sub-cell under short circuit condition

The photo-generated carriers (holes) in the c-Si absorber region reach the junction through diffusion and get collected at the front contact. The a-Si layers at the back of the device act as effective back surface field and avoids recombination at the wrong contact. The properties of the a-Si/c-Si heterojunction (see Fig. S10) such as (band offset, emitter doping, etc.) at the junction dictate the performance of the device.

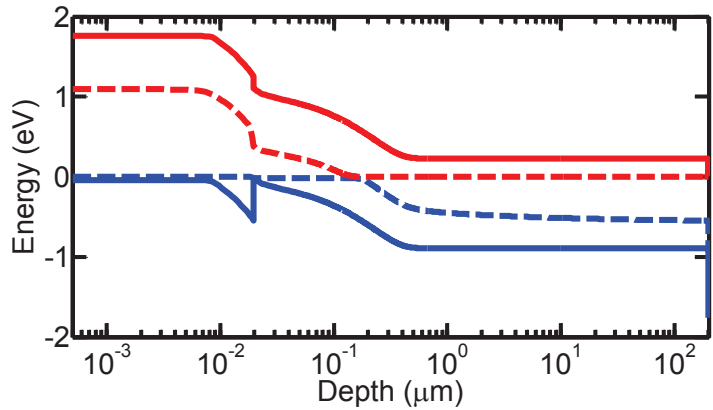

**Fig. S10** The energy band diagram of the HIT sub-cell under short circuit condition

## Fabrication prospects for bifacial tandem configuration:

As observed from this study, the bifacial tandem cell structure shows prospects of improved performance with small changes in the existing experimental setups. Here we discuss some of the minor modifications needed in current deposition methods to achieve this goal. The bottom HIT sub-cell can be grown with conventional deposition technologies.[6] PCBM, Perovskite, and PEDOT:PSS layers can be grown using solution process techniques—these use organic solvents and low temperatures. Thus the processing of the Perovskite sub-cell is unlikely to damage the bottom sub-cell. However, the deposition of ITO using sputtering can be challenging and may damage the organic layers. Fortunately, it is possible to use other transparent electrodes such as Ag-NW nets,[7] or graphene-NW co-percolating networks.[8] Some of these electrodes have been demonstrated to work on organic photovoltaics and Perovskite cells.[9,10] Although the bifacial Perovskite-HIT tandem seems promising, the exact structure and processing steps would require further research.

**Table S2.** The simulation parameters used in numerical simulation of HIT cell[11]

| Properties | p-layer | i-layer | c-Si | i-layer | n-layer |
|---|---|---|---|---|---|
| Thickness | 10nm | 10nm | 200μm | 10nm | 10nm |
| Doping ($cm^{-3}$) | $N_A = 5 \times 10^{19}$ | – | $N_D = 5 \times 10^{15}$ | – | $N_D = 5 \times 10^{19}$ |
| Hole mobility ($cm^2/Vs$) | 0.2 | 2 | 330 | 2 | 2 |
| Electron mobility ($cm^2/Vs$) | 1 | 20 | 1030 | 20 | 10 |
| Hole lifetime (s) | – | – | $5 \times 10^{-4}$ | – | – |
| Electron lifetime (s) | – | – | $5 \times 10^{-4}$ | – | – |
| Band gap ($eV$) | 1.7 | 1.7 | 1.12 | 1.7 | 1.7 |
| Electron affinities ($eV$) | 3.9 | 3.9 | 4.05 | 3.9 | 3.9 |
| Conduction Band Tail States | $N_{tail} = 10^{19} cm^{-3} eV^{-1}$<br>$E_{tail} = 0.05 eV$<br>$C_h = 10^{-16} cm^{-2}$<br>$C_n = 10^{-16} cm^{-2}$ | $N_{tail} = 10^{19} cm^{-3} eV^{-1}$<br>$E_{tail} = 0.019 eV$<br>$C_h = 10^{-16} cm^{-2}$<br>$C_n = 10^{-16} cm^{-2}$ | – | $N_{tail} = 10^{19} cm^{-3} eV^{-1}$<br>$E_{tail} = 0.019 eV$<br>$C_h = 10^{-16} cm^{-2}$<br>$C_n = 10^{-16} cm^{-2}$ | $N_{tail} = 10^{19} cm^{-3} eV^{-1}$<br>$E_{tail} = 0.05 eV$<br>$C_h = 10^{-16} cm^{-2}$<br>$C_n = 10^{-16} cm^{-2}$ |
| Valence band tail states | $N_{tail} = 10^{19} cm^{-3} eV^{-1}$<br>$E_{tail} = 0.1 eV$<br>$C_h = 10^{-16} cm^{-2}$<br>$C_n = 10^{-16} cm^{-2}$ | $N_{tail} = 10^{19} cm^{-3} eV^{-1}$<br>$E_{tail} = 0.049 eV$<br>$C_h = 10^{-16} cm^{-2}$<br>$C_n = 10^{-16} cm^{-2}$ | – | $N_{tail} = 10^{19} cm^{-3} eV^{-1}$<br>$E_{tail} = 0.049 eV$<br>$C_h = 10^{-16} cm^{-2}$<br>$C_n = 10^{-16} cm^{-2}$ | $N_{tail} = 10^{19} cm^{-3} eV^{-1}$<br>$E_{tail} = 0.1 eV$<br>$C_h = 10^{-16} cm^{-2}$<br>$C_n = 10^{-16} cm^{-2}$ |
| Donor-like defects | $N_{t1} = 10^{16} cm^{-3}$<br>$E_{t1} - E_i = 0.31\ eV$<br>$\sigma_{t1} = 0.08\ eV$<br>$\tau_{p1} = 10^{-9} s$<br>$\tau_{n1} = 10^{-9} s$<br>$N_{t2} = 10^{18} cm^{-3}$<br>$E_{t2} - E_i = -0.26\ eV$<br>$\sigma_{t2} = 0.15\ eV$<br>$\tau_{p2} = 10^{-9} s$<br>$\tau_{n2} = 10^{-9} s$ | $N_{t1} = 5 \times 10^{15} cm^{-3}$<br>$E_{t1} - E_i = 0.21\ eV$<br>$\sigma_{t1} = 0.08\ eV$<br>$\tau_{p1} = 2 \times 10^{-9} s$<br>$\tau_{n1} = 2 \times 10^{-9} s$ | – | $N_{t1} = 5 \times 10^{15} cm^{-3}$<br>$E_{t1} - E_i = 0.21\ eV$<br>$\sigma_{t1} = 0.08\ eV$<br>$\tau_{p1} = 2 \times 10^{-9} s$<br>$\tau_{n1} = 2 \times 10^{-9} s$ | $N_{t1} = 10^{16} cm^{-3}$<br>$E_{t1} - E_i = 0.31\ eV$<br>$\sigma_{t1} = 0.08\ eV$<br>$\tau_{p1} = 10^{-9} s$<br>$\tau_{n1} = 10^{-9} s$<br>$N_{t2} = 10^{18} cm^{-3}$<br>$E_{t2} - E_i = -0.26\ eV$<br>$\sigma_{t2} = 0.15\ eV$<br>$\tau_{p2} = 10^{-9} s$<br>$\tau_{n2} = 10^{-9} s$ |
| Acceptor-like defects | $N_{t1} = 10^{16} cm^{-3}$<br>$E_{t1} - E_i = -0.21\ eV$<br>$\sigma_{t1} = 0.08\ eV$<br>$\tau_{p1} = 10^{-9} s$<br>$\tau_{n1} = 10^{-9} s$ | $N_{t1} = 5 \times 10^{15} cm^{-3}$<br>$E_{t1} - E_i = 0.11\ eV$<br>$\sigma_{t1} = 0.08\ eV$<br>$\tau_{p1} = 2 \times 10^{-9} s$<br>$\tau_{n1} = 2 \times 10^{-9} s$ | – | $N_{t1} = 5 \times 10^{15} cm^{-3}$<br>$E_{t1} - E_i = 0.11\ eV$<br>$\sigma_{t1} = 0.08\ eV$<br>$\tau_{p1} = 2 \times 10^{-9} s$<br>$\tau_{n1} = 2 \times 10^{-9} s$ | $N_{t1} = 10^{16} cm^{-3}$<br>$E_{t1} - E_i = -0.21\ eV$<br>$\sigma_{t1} = 0.08\ eV$<br>$\tau_{p1} = 10^{-9} s$<br>$\tau_{n1} = 10^{-9} s$ |
| Contact properties | Ohmic contacts, $s_f = 10^7 cm/s$ for both contacts | | | | |

**Table S3.** The simulation parameters used in numerical simulation of Perovskite cell

| Properties | PEDOT:PSS | Perovskite | PCBM |
|---|---|---|---|
| Thickness | 40nm | 135nm and 350nm | 20nm |
| Doping ($cm^{-3}$) | $N_A = 3 \times 10^{17}$ | $N_A = 1 \times 10^{16}$ | $N_D = 5 \times 10^{17}$ [12] |
| Hole mobility ($cm^2/Vs$) | $9 \times 10^{-3}$ [13] | 40 | $1 \times 10^{-2}$ |
| Electron mobility ($cm^2/Vs$) | $9 \times 10^{-3}$ | 40 | $1 \times 10^{-2}$ [12] |
| Hole lifetime (s) | $1 \times 10^{-6}$ | $20 \times 10^9$ | $1 \times 10^{-6}$ |
| Electron lifetime (s) | $1 \times 10^{-6}$ | $20 \times 10^9$ | $1 \times 10^{-6}$ |
| Band gap ($eV$) | 1.55 [14] | 1.55 [15] | 2.0 [16] |
| Electron affinities ($eV$) | 3.63 | 3.73 [17] | 4.17 [18] |
| Contact Properties | Ohmic contacts, $s_f = 10^7 cm/s$ for both contacts | | |

**References:**

[1] G. Xing, N. Mathews, S.S. Lim, N. Yantara, X. Liu, D. Sabba, M. Grätzel, S. Mhaisalkar, and T.C. Sum, Nat. Mater. **13**, 476 (2014).
[2] J.M. Ball, S.D. Stranks, M.T. Hörantner, S. Hüttner, W. Zhang, E.J.W. Crossland, I. Ramirez, M. Riede, M.B. Johnston, R.H. Friend, and H.J. Snaith, Energy Environ. Sci. **8**, 602 (2015).
[3] P. Löper, M. Stuckelberger, B. Niesen, J. Werner, M. Filipič, S.-J. Moon, J.-H. Yum, M. Topič, S. De Wolf, and C. Ballif, J. Phys. Chem. Lett. **6**, 66 (2014).
[4] M. Filipič, P. Löper, B. Niesen, S. De Wolf, J. Krč, C. Ballif, and M. Topič, Opt. Express **23**, A263 (2015).
[5] W. Nie, H. Tsai, R. Asadpour, J.-C. Blancon, A.J. Neukirch, G. Gupta, J.J. Crochet, M. Chhowalla, S. Tretiak, M.A. Alam, H.-L. Wang, and A.D. Mohite, Science **347**, 522 (2015).
[6] C. Ballif, S. De Wolf, A. Descoeudres, and Z.C. Holman, *Advances in Photovoltaics: Part 3* (Elsevier, 2014), pp. 73–120.
[7] C.D. Bailie, M.G. Christoforo, J.P. Mailoa, A.R. Bowring, E.L. Unger, W.H. Nguyen, J. Burschka, N. Pellet, J.Z. Lee, M. Grätzel, R. Noufi, T. Buonassisi, A. Salleo, and M.D. McGehee, Energy Environ. Sci. **8**, 956 (2015).
[8] R. Chen, S.R. Das, C. Jeong, M.R. Khan, D.B. Janes, and M.A. Alam, Adv. Funct. Mater. (2013).
[9] R. Zhu, C.-H. Chung, K.C. Cha, W. Yang, Y.B. Zheng, H. Zhou, T.-B. Song, C.-C. Chen, P.S. Weiss, G. Li, and Y. Yang, ACS Nano **5**, 9877 (2011).
[10] W. Gaynor, G.F. Burkhard, M.D. McGehee, and P. Peumans, Adv. Mater. **23**, 2905 (2011).
[11] R.V.K. Chavali, S. Khatavkar, C.V. Kannan, V. Kumar, P.R. Nair, J.L. Gray, and M.A. Alam, IEEE J. Photovolt. **5**, 725 (2015).
[12] V.A. Trukhanov, V.V. Bruevich, and D.Y. Paraschuk, Phys. Rev. B **84**, 205318 (2011).
[13] S.A. Rutledge and A.S. Helmy, J. Appl. Phys. **114**, 133708 (2013).
[14] A. Lenz, H. Kariis, A. Pohl, P. Persson, and L. Ojamäe, Chem. Phys. **384**, 44 (2011).
[15] E. Mosconi, A. Amat, M.K. Nazeeruddin, M. Grätzel, and F. De Angelis, J. Phys. Chem. C **117**, 13902 (2013).

[16] Z.-L. Guan, J.B. Kim, H. Wang, C. Jaye, D.A. Fischer, Y.-L. Loo, and A. Kahn, Org. Electron. **11**, 1779 (2010).
[17] P. Schulz, E. Edri, S. Kirmayer, G. Hodes, D. Cahen, and A. Kahn, Energy Environ. Sci. **7**, 1377 (2014).
[18] B.W. Larson, J.B. Whitaker, X.-B. Wang, A.A. Popov, G. Rumbles, N. Kopidakis, S.H. Strauss, and O.V. Boltalina, J. Phys. Chem. C **117**, 14958 (2013).